\documentclass[reprint,amsmath,amssymb,aps,prb]{revtex4-2}
\usepackage[utf8]{inputenc}
\usepackage{dcolumn}
\usepackage{bm}
\usepackage{graphics}
\usepackage[pdftex]{graphicx}
\usepackage[export]{adjustbox}
\usepackage{pgfplots,pgfplotstable}
\usepackage{tikz}
\pgfplotsset{compat=newest}

\begin{document}
\title{Dissipation by surface states in superconducting RF cavities}
\author{Sean Deyo}
\email{sjd257@cornell.edu}
\author{Michelle Kelley}
\author{Nathan Sitaraman}
\author{Danilo B. Liarte}
\author{Tomas Arias}
\author{James P. Sethna}
\affiliation{Laboratory of Atomic and Solid State Physics, Cornell University, Ithaca, NY, USA}
\author{Thomas Oseroff}
\author{Matthias Liepe}
\affiliation{Cornell Laboratory for Accelerator-Based Sciences and Education, Cornell University, Ithaca, NY, USA}
\date{\today}

\begin{abstract}
Recent experiments on superconducting cavities have found that under large radio-frequency (RF) electromagnetic fields the quality factor can improve with increasing field amplitude, a so-called ``anti-$Q$ slope.'' Linear theories of dissipation break down under these extreme conditions and are unable to explain this behavior. We numerically solve the Bogoliubov-de Gennes equations at the surface of a superconductor in a parallel AC magnetic field, finding that at large fields there are quasiparticle surface states with energies below the bulk value of the superconducting gap. As the field oscillates, such states emerge and disappear with every cycle. We consider the dissipation resulting from inelastic quasiparticle-phonon scattering into these states and investigate the ability of this mechanism to explain features of the experimental observations, including the field dependence of the quality factor. We find that this mechanism is likely not the dominant source of dissipation and does not produce an anti-$Q$ slope by itself; however, we demonstrate in a modified two-fluid model how these bound states can play a role in producing an anti-$Q$ slope.
\end{abstract}

\maketitle

\section{Introduction}
\label{sec:intro}
Superconducting radio-frequency (SRF) cavities are useful in a variety of modern applications, such as free-electron lasers and particle colliders \cite{Dhakal}. The primary advantage of SRF cavities over their normal-conducting alternatives is a lower surface resistance, which has the obvious benefit of a lower energy footprint \cite{Padamsee}. Despite decades of steady improvements, allowing for quality factors in the neighborhood of $10^{10}$ \cite{Dhakal} and accelerating fields up to $25-45$ MV/m \cite{Padamsee}, some unanswered questions remain. At high fields there is often a ``$Q$ slope,'' meaning the quality factor decreases---or, equivalently, surface resistance increases---as field strength increases. More confounding, recently some cavities, in particular niobium cavities doped with nitrogen, have exhibited an ``anti-$Q$ slope,'' meaning that the quality factor actually increases with field \cite{Dhakal}. Increasing the resonant frequency of the cavity has been associated with stronger anti-$Q$ slopes \cite{Martinello}. Since inelastic scattering rates \cite{Kaplan} can be comparable to typical SRF frequencies \cite{Gonnella}, this suggests inelastic scattering may play an important role in determining the field dependence of surface resistance. 

Definitive theoretical explanations for the anti-$Q$ slope remain elusive.
Conventional theories of AC dissipation in superconductors predict no $Q$ slope (constant quality factor).
Mattis and Bardeen~\cite{MattisBar1958} employed linear response methods along with BCS theory~\cite{BardeenSch1957} to calculate the complex conductivity of a superconductor subject to a magnetic field.
They found a surface resistance that roughly goes as $\omega^2\exp (-\Delta / k_B T)$ at low temperatures, where $\Delta$ is the superconducting gap and $\omega$ is the angular frequency of the field.
Central to these calculations is the idea that a quasiparticle in a state with energy $E_1$ can absorb a photon of energy $\hbar\omega$, and transition to a state of energy $E_2=E_1+\hbar\omega$. 
(Here we ignore other sources of dissipation such as trapped magnetic vortices oscillating near the cavity surface~\cite{GurevichCio2013,LiarteSet2018,ChecchinSer2018} and within grain boundaries~\cite{SheikhzadaGur2017,CarlsonPos2021}.)
In order for the photon energy to be well defined the absorption must happen coherently over many cycles of the AC perturbation.

Extensions of the linear-response theory of~\cite{MattisBar1958} have been routinely invoked to interpret experimental data showing an anti-$Q$ slope. 
Gurevich applied the Keldysh formalism in the context of non-equilibrium Green's functions to rederive the conductivity and evaluate the surface resistance at low frequencies and mean free paths, and high magnetic fields~\cite{Gurevich}.
The anti-$Q$ slope there originates in the smearing of the density of states due to oscillating superflow.
At high enough fields, the decrease of the superconducting gap and subsequent smearing of the density of states weakens the zero-frequency singularity of the Mattis-Bardeen resistance, leading to a field dependence.
This calculation involves an approximation to determine the nonequilibrium quasiparticle distribution function, a key quantity that is very difficult to find starting from a fundamental theory.
Goldie and Withington obtained non-linear solutions of the kinetic equations for the coupled quasiparticle and phonon systems~\cite{GoldieWit2012}.
Their solutions for the non-thermal quasiparticle distribution function were later combined with the Mattis-Bardeen theory by de Visser et al., who proposed a mechanism for microwave suppression on superconducting aluminum resonators~\cite{VisserKla2014}.

Though some of these extensions can be used in the regimes of strong fields~\cite{Gurevich,Kubo}, they maintain the notion of $E_1\to E_2=E_1+\hbar\omega$. At large fields, however, we shall see that there are quasiparticle surface states whose  energies change dramatically during each cycle. The situation is reminiscent of the smeared density of states in \cite{Gurevich}, except that we explicitly include depth and time dependence of both the superconducting gap and the quasiparticle states. In particular, we find that $E_2(t)-E_1(t)$ is not constant for such states, making the equality $E_2=E_1+\hbar\omega$ questionable. This discrepancy demands a new framework to handle such states in strong AC fields. Rather than consider transitions occurring coherently over many cycles, we take the opposite limit and consider scattering events occurring within a single cycle: solving the same quantum dissipation problem, but starting from the adiabatic limit rather than the weak-coupling limit. 

A simpler phenomenological model relying on the two-fluid idea~\cite{Halbritter1974} yields explicit analytic formulas for the surface resistance but lacks essential physical ingredients present in the Mattis-Bardeen calculation, such as coherent effects on the transition probabilities~\cite{Tinkham}. We briefly discuss how to incorporate the quasiparticle states with changing energies into a modified version of the two-fluid model.

In Section~\ref{sec:formalism} we introduce the Bogoliubov-de Gennes equations and describe quasiparticle surface states whose energies exhibit strong field dependence. In Section~\ref{sec:mechanism} we consider how quasiparticle-phonon scattering changes the occupation of these surface states during the RF cycle and can lead to dissipation. In Section~\ref{sec:results} we compute the field and frequency dependence of our dissipation mechanism and discuss its relevance to the anti-$Q$ slope. Finally, in Section~\ref{sec:conclusion} we offer concluding remarks and possibilities for further study.

\section{Bogoliubov-de Gennes formalism}
\label{sec:formalism}
To find quasiparticle states we solve the Bogoliubov-de Gennes {(BdG)} equations \cite{deGennes}:
\begin{equation}
\begin{aligned}
(H_e+U)u+\Delta v&=Eu\\
-(H_e^*+U)v+\Delta^* u&=Ev
\end{aligned}
\label{eq:BdG}
\end{equation}
with
\begin{equation*}
H_e(\mathbf{r})=(-i\hbar\nabla-e\mathbf{A}(\mathbf{r})/c)^2/2m+U_0(\mathbf{r})-E_F,
\end{equation*}
where $e$ is the fundamental charge, $c$ is the speed of light, and $m$ is the mass of an electron. These equations must be solved self-consistently with the potentials given by
\begin{equation}
\begin{aligned}
U&=-V\sum_n |u_n|^2 f_n + |v_n|^2 (1-f_n)\\
\Delta&=V\sum_n u_nv_n^*(1-2f_n),
\label{eq:potentials}
\end{aligned}
\end{equation}
where $f_n$ is the occupation of state $n$ and $V$ is a constant describing the strength of the interaction. To accomplish this numerically, we make an initial guess for the potentials, find all the solutions $(u_n,v_n)$, use them to compute a refined guess for the potentials, and iterate until the process settles on a set of $(u_n,v_n)$ consistent with $U$ and $\Delta$. Note that Eqns.~\eqref{eq:BdG} possess a symmetry: If $(u,v)$ is a solution with energy $E$, $(v^*,-u^*)$ is a solution with energy $-E$. If one takes advantage of this symmetry one must also replace the state's occupation $f$ with $1-f$ in order to preserve Eqns.~\eqref{eq:potentials}. In doing so, one treats the quasiparticle state as a `quasi-hole' state with opposite energy and occupation.

We solve Eqns.~\eqref{eq:BdG} at the surface of a superconductor in a parallel magnetic field: $\mathbf{A}=A_0\sin(\omega t)e^{-z/\lambda}\hat{y}$, where $\lambda$ is the London penetration depth of the superconductor. {For simplicity we impose the vector potential externally, rather than using the current density and Maxwell's equations to solve for a new vector potential after each iteration of the self-consistency process. However, we can confirm that our current density, after reaching self-consistency with $U$ and $\Delta$, is also consistent with our vector potential. This accords with observations that $\lambda$ does not change significantly for RF fields below $H_{c1}$ \cite{Carlson}.}

For the sake of simplicity and generality we neglect the atomic potential $U_0$ and assume a spherical Fermi surface. To avoid simulating a semi-infinite half-space, we limit the domain to a depth $L$. Provided $L\gg\lambda$, this truncation does not affect the energies of states localized near the surface. We also take the occupation fractions to be in equilibrium, $f_n=\left(1+e^{\beta E_n}\right)^{-1}$, when self-consistently finding the potentials, having verified numerically that this approximation is of little effect.

Factoring out the dependence in the directions parallel to the surface, $u(\mathbf{r})\to e^{i(k_xx+k_yy)}u(z)$ and likewise for $v$, we obtain a coupled system
of differential equations: \begin{widetext}
\begin{equation}
\begin{aligned}
\left\{\frac{\hbar^2}{2m}\left[k_x^2+\left(k_y-\frac{eA_0}{\hbar c}\sin(\omega t)e^{-z/\lambda}\right)^2-\frac{d}{dz}^2\right]-E_F+U(z)\right\}u(z)+\Delta(z) v(z)&=Eu(z),\\
-\left\{\frac{\hbar^2}{2m}\left[k_x^2+\left(k_y+\frac{eA_0}{\hbar c}\sin(\omega t)e^{-z/\lambda}\right)^2-\frac{d}{dz}^2\right]-E_F+U(z)\right\}v(z)+\Delta(z) u(z)&=Ev(z),
\end{aligned}
\label{eq:odes}
\end{equation}
\end{widetext}
with $k_x$ and $k_y$ as parameters. Since the time dependence of the field $A_0(t)$ is much slower than quantum relaxation times (apart from inelastic scattering), we obtain the time dependence of the eigenstates simply by solving these equations at a series of times.

For boundary conditions we set $u(0)=u(L)=v(0)=v(L)=0$ to confine the quasiparticles to the slab. One can see from Eqns.~\eqref{eq:potentials} that this forces the potentials to vanish at the surface, which might seem to be a problem given that, for instance, $\Delta$ should be a nonzero constant if the field is zero. However, the relevant length scale for features of $\Delta$ is the {correlation length}, while the sums in Eqns.~\eqref{eq:potentials} include terms that oscillate on the much shorter length scale of $1/k_F$. For niobium $1/k_F=0.08$ nm \cite{Ashcroft}, about $500$ times smaller {than the correlation length of nearly $40$ nm \cite{Gonnella}}. Thus, {any oscillations due to} forcing $\Delta(0)=0$ are only present within a tiny fraction of a penetration depth from the surface, reminiscent of the Friedel oscillations in electron density near a surface \cite{Crommie}. Others who have studied the Bogoliubov-de Gennes equations near a surface observe the same phenomenon \cite{Troy}, or a similar phenomenon for the BdG equations with the tight-binding model \cite{Croitoru}. {Authors studying superconductivity in thin films and wires have also noted oscillations in the energy gap as a function of thickness \cite{blatt1963shape,shanenko2006shape}. It is important to distinguish between oscillations in the energy gap as a function of thickness in a thin film and oscillations in the pair potential as a function of position in a thick slab or half-space, but both effects ultimately result from boundary conditions. Alternative boundary conditions, as for instance in the jellium model \cite{Lang}, could soften the oscillations in the pair potential, but we would not expect them to disappear entirely.}

\section{Dissipation mechanism}
\label{sec:mechanism}
\begin{figure}[t]
\includegraphics[width=.48\textwidth,left]{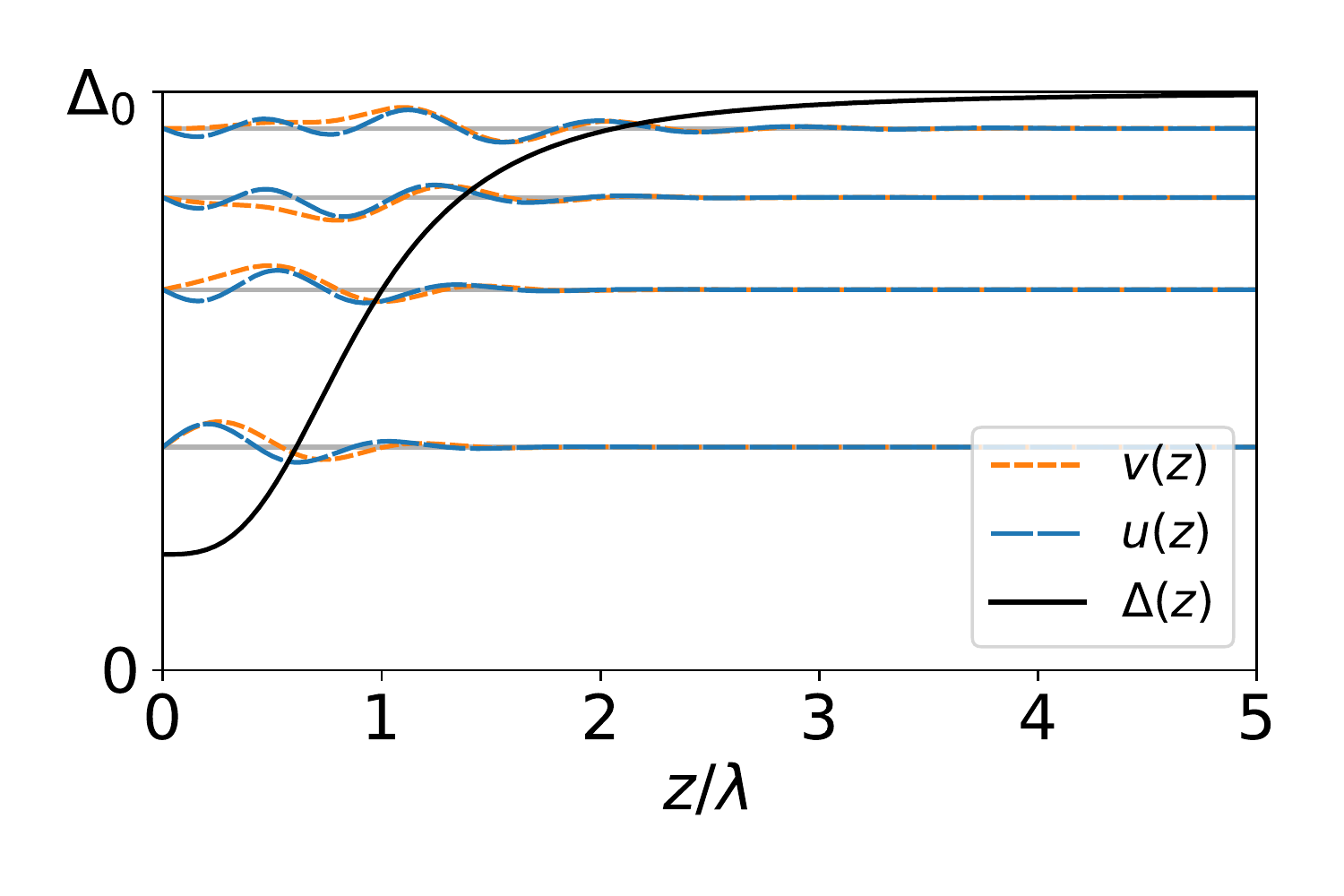}
\caption{The BdG quasiparticle bound states for $k_x=k_F$, $k_y=0$, at a field of $40$ MV/m. For each state the wave functions $u$ and $v$ are plotted with a vertical shift according to their energy. The situation is somewhat analogous to a particle in a potential $\Delta(z)$ with a hard wall on the left.}
\label{fig:bd}
\end{figure}

In the absence of a field, there is a uniform pair potential $\Delta(z)=\Delta_0$ and all quasiparticle energies must be larger than $\Delta_0$. Once the field is applied, states with energies less than $\Delta_0$ appear.
These states are localized near the surface. Examples are plotted in Figure~\ref{fig:bd}.
There is also a continuum of states, not localized to the surface, whose energies change negligibly during the cycle. Because elastic scattering is quick---mean-free path estimates suggest an elastic scattering time in the femtosecond range, several orders of magnitude quicker than typical SRF field frequencies in the GHz range \cite{Liepe}---we can assume that a state stays in equilibrium as long as it can scatter elastically with this reservoir of continuum states. Bound states with energies below $\Delta_0$ cannot scatter elastically with continuum states, so they can only be filled by the much slower \textit{in}elastic processes. If the inelastic scattering rate is comparable to or slower than $\omega$ there will be disequilibrium. 

The fact that the scattering rate can change during the cycle complicates the dynamics. For any fermion state it is true that
\begin{equation}
    \frac{df}{dt} = \frac{1-f}{\tau_h} - \frac{f}{\tau_p}
    \label{eq:dfdt}
\end{equation}
where $\tau_h$ and $\tau_p$ are the hole and particle lifetimes, respectively. Setting Eqn.~\eqref{eq:dfdt} to zero allows us to relate the equilibrium occupation $f_0$ to the two lifetimes:
\begin{equation}
    (1-f_0)\tau_p = f_0 \tau_h \qquad \text{or} \qquad f_0 = \frac{\tau_p}{\tau_p+\tau_h}.
    \label{eq:f0}
\end{equation}
We can use these relations to eliminate either of the two lifetimes:
\begin{equation}
    \frac{df}{dt} = \frac{f_0-f}{f_0\tau_h} = \frac{f_0-f}{(1-f_0)\tau_p}.
    \label{eq:ft}
\end{equation}
At least one of the lifetimes must be time-dependent if the energy is time-dependent, since $f_0=(1+e^{\beta E})^{-1}$. The simplest approximation is to take one of the lifetimes to be constant and let the other {be determined by} Eqn.~\eqref{eq:f0}. For a bound state at energy $E<\Delta_0$, $\tau_p$ is determined in large part by the ability to absorb a phonon of energy $\Delta_0-E$. The occupation numbers for such phonons depend exponentially on $\Delta_0-E$, making $\tau_p$ strongly time-dependent. $\tau_h$, on the other hand, is determined largely by spontaneous emission of phonons, making the occupation numbers irrelevant. Thus, constant $\tau_h$ is the more physical choice for a first approximation. Measurements in bulk niobium indicate an inelastic scattering lifetime of $\tau_p=16$ ns at $2$ K for a quasiparticle at $E=\Delta_0$ \cite{Kaplan}, which corresponds to $\tau_h=100$ µs.

{To refine the approximation, let us consider what might} make the hole lifetime not constant. There is a simple phase space argument for the energy dependence of $\tau_h$ in the low-temperature regime: Kaplan et al. provide well-known formulas for the inelastic scattering rates of quasiparticles in terms of the Eliashberg spectral function, $\alpha^2 F(\Omega)$, which is approximately quadratic in the phonon energy $\Omega$ at low frequencies~\cite{Kaplan}. Consider one quasiparticle bound state with energy $E(t)<\Delta_0$: the typical scattering processes at low temperatures are a quasiparticle in a continuum state with energy $\Delta_0+k_BT$ emitting a phonon and entering the bound state, and a quasiparticle in the bound state absorbing a phonon and entering a continuum state with energy $\Delta_0+k_BT$. These rates are proportional to
\begin{equation*}
\alpha^2 F(\Omega) \propto \Omega^2 \propto (\Delta_0+k_BT-E(t))^2.
\end{equation*}
Furthermore, because bound state wave functions are concentrated within one penetration depth of the surface, the scattering rate between a bound state and a continuum state is roughly proportional to the local density of the continuum state at the surface. We can compute the average density of continuum quasiparticles within one penetration depth of the surface, $\rho_1$, as a function of field and compare it to the average density at zero field, $\rho_0$. {Combining the phase space argument with the surface density effect, we conclude that the scattering rate is approximately proportional to $\alpha^2 F(\Omega) \cdot \rho_1/\rho_0$. Thus, our refined approximation for the hole lifetime is}
\begin{equation}
\tau_h(t) = \tau_0 \frac{\rho_0}{\rho_1(t)} \left(\frac{k_BT_c}{\Delta_0+k_BT-E(t)}\right)^2,
\label{eq:tauh}
\end{equation}
where $T_c$ is the superconducting critical temperature and the constant $\tau_0$ is set by demanding $\tau_h=100$~µs at $2$~K for a state at $E=\Delta_0$.

{Even with these refinements, our model of $\tau_h$ is still an approximation. We have not explicitly considered possible effects of surface ordering in enhancing $T_c$ at the surface \cite{ginzburg1964surface}. This could be an important point to consider for a more detailed model. Merely changing $T_c$ in Eqn. \eqref{eq:tauh} would have little effect, given that $T_c$ only appears as a multiplicative constant and the coefficient $\tau_0$ fixes the overall scaling of $\tau_h$; however, the physics of surface ordering (among other things) could demand a depth-dependent expression for $\tau_h$.}

With these scattering rates we can calculate $f(t)$. Every time the field goes to zero the bound states return to the continuum, where they scatter elastically with continuum states and return almost instantaneously to $f=f_0$, dumping their excess occupation into the reservoir of continuum states. The process repeats when the field starts to increase again and bound states reemerge. Thus, for each bound state we can find the net work done by the field in a single half cycle and multiply by $\omega/\pi$ to obtain the power dissipated: 
\begin{equation}
P_\text{diss}=\frac{\omega}{\pi}\int_0^{T/2} f(t)\frac{dE}{dt} dt.
\end{equation}
In this form it is clear that the dissipation comes from the changing energies of occupied states. Ultimately the excess energy is dumped via phonons, and to illustrate this one can integrate by parts to obtain an equivalent formula:
\begin{equation}
P_\text{diss}=\frac{\omega}{\pi}\int_0^{T/2} \frac{df}{dt}\left(E_0-E(t)\right) dt.
\end{equation}
The rate of phonon emission is $df/dt$, and the phonon energy is $E_0-E(t)$. The formula holds for any $E_0$; that is, any energy the quasiparticle may have come from before emitting the phonon. When using this formula one must be careful to handle the delta function in $df/dt$ that occurs when the bound state returns to the continuum at the end of the half cycle.

At large fields some bound states reach negative energies. When this happens one can use the `quasi-hole' symmetry described in Section~\ref{sec:formalism}. If the quasiparticle energy drops below $-\Delta_0$ or, equivalently, the quasi-hole reaches an energy above $\Delta_0$, the state can briefly scatter elastically with continuum states until it reenters the interval $(-\Delta_0,\Delta_0)$.

\begin{figure}
\includegraphics[width=.45\textwidth,left]{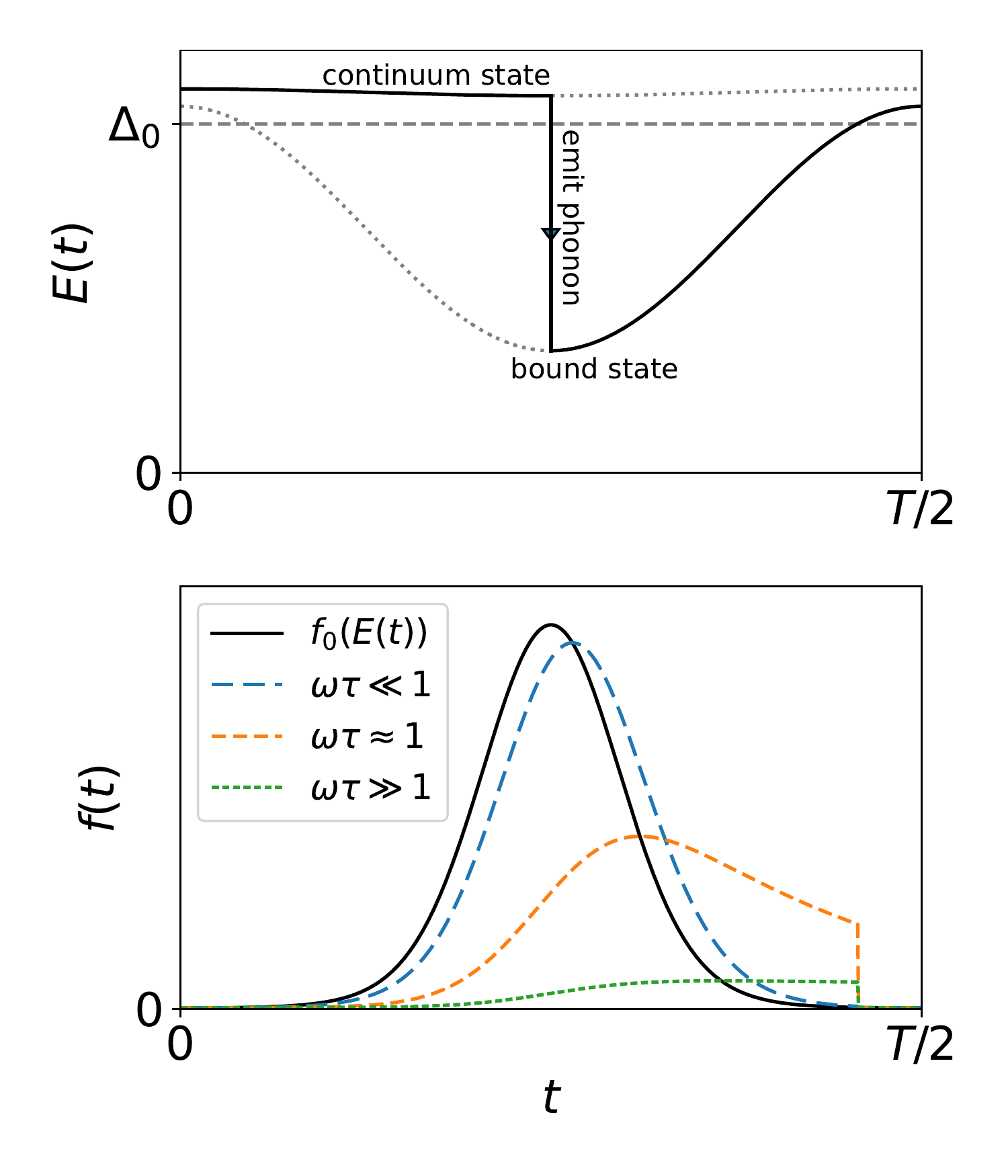}
\caption{Top: quasiparticle scattering from a bulk state into a bound state, releasing a phonon in the process. As the bound state energy rises, the field does work on the quasiparticle, resulting in net dissipation. Bottom: the macroscopic picture tracking the average occupation of an ensemble of such states for several values of $\omega\tau$. Dissipation is maximized when ${\omega\tau\approx1}$.}
\label{fig:ft}
\end{figure}

Figure~\ref{fig:ft} illustrates the dynamics. For simplicity, the figure was made assuming ${df/dt=-(f-f_0)/\tau}$, which is the low-field limit of Eqn.~\eqref{eq:ft} with $\tau=\tau_p$. In this limit we can explicitly solve for $f(t)$: 
\begin{equation*}
f(t)=e^{-t/\tau}\left[f_0(E(0)) + \frac1\tau \int_0^{t} e^{s/\tau} f_0(E(s)) ds\right].
\end{equation*}
One can see that $f$ is larger in the second half of the plot because it lags behind $f_0$. The second half is when $dE/dt$ is positive, so this imbalance is the reason the field does positive net work. The size of the imbalance depends on $\tau$: If inelastic scattering is much quicker than the field (${\omega\tau\ll1}$), the state stays near equilibrium the whole time ($f$ is close to $f_0$) and little dissipation occurs. If inelastic scattering is slow (${\omega\tau\gg1}$), the state hardly ever gets filled ($f$ barely changes) and again little dissipation occurs. The maximum imbalance occurs for ${\omega\tau\approx1}$. At typical SRF frequencies of a few GHz, niobium's ${\tau_p=16\text{ ns}}$ puts us in the ${\omega\tau\gg 1}$ regime.

We stress that this mechanism does not apply to unbound states, whose wave functions oscillate throughout the slab and are not concentrated near the surface. We discuss the dissipation due to these states in Section~\ref{sec:results}, using a modified two-fluid formalism~\cite{Tinkham}.

After computing the dissipation over an infinitesimal area $ds$, we find the surface resistance $R_\text{s}$ from 
\begin{equation}
P_{\text{diss}}=\frac12 R_\text{s} \int|H|^2 \, ds
\end{equation}
and the quality factor from
\begin{equation}
Q=\frac{\omega U}{P_{\text{diss}}}=\frac{\omega\mu_0\int |H|^2 \, dv}{R_\text{s}\int |H|^2 \, ds}=\frac{G}{R_\text{s}}
\end{equation}
where $G=\frac{\omega\mu_0\int |H|^2 \, dv}{\int |H|^2 \, ds}$ is the geometry factor, a parameter independent of the frequency and size of the SRF cavity. In order to, say, double the frequency of a cylindrical cavity, one must halve the radius of the cavity in order to keep it in the same wave guide mode, so $G$ remains unchanged and is in fact purely geometrical.

We use parameters realistic for an elliptical SRF cavity made from niobium: a Fermi energy of $E_F=5.32$~eV \cite{Ashcroft}, a bulk gap of $\Delta_0=1.5$~meV, a penetration depth of $\lambda=40$~nm, a temperature of $2$~K, a geometry factor of $G=270$~$\Omega$, a frequency of $1.3$~GHz, and accelerating gradients up to $30$~MV/m \cite{Gonnella}. Such gradients correspond to magnetic fields of more than half of niobium's lower critical field $H_{c1}$ at $2$ K \cite{French}.

Additionally, we can use the Bogoliubov-de Gennes states we calculate as inputs into the two-fluid model{, with superconducting electrons treated} as an inductive channel and normal conducting electrons as a resistive channel. With these two channels in parallel, and in the limit that the inductive channel has a much lower impedance, the AC dissipation is proportional to the density of the normal fluid:
\begin{equation}
R_\text{s} \approx \delta^{-1}\sigma_1/\sigma_2^2
\label{eq:2fl}
\end{equation}
with
\begin{equation}
\begin{aligned}
\sigma_1 &= n_n e^2 \tau_n / m \\
\sigma_2 &= n_s e^2 / m \omega
\end{aligned}
\end{equation}
where $\delta\approx c\,(4\pi\omega\sigma_2)^{-1/2}$ is the skin depth, $n_n$ and $n_s$ are the densities of the normal and superconducting electrons, and $\tau_n$ is the relaxation time of the normal electrons~\cite{Tinkham}. Note that $\tau_n$ is not the same as the inelastic scattering time; it is a phenomenological parameter one can calculate from the electrical resistivity, which tends to be dominated by elastic scattering with impurities at low temperatures~\cite{Ashcroft,Webb}. 

{To estimate the local fraction of normal electrons in the superconducting phase, we compute the proportion of electrons relative to the normal phase available for conduction
\begin{equation}
\frac{n_n}{n}(z)=\frac{1}{N(0)}\sum_n\left( -\frac{\partial f_n}{\partial E_n}\right) \left(u_n^2(z)+v_n^2(z)\right),
\label{eq:frac}
\end{equation}
where $N(0)$ denotes the electronic density of states at the Fermi level. This quantity, and other normal-phase properties, are readily computed within the Bogoliubov-de Gennes framework by forcing $\Delta=0$ in Eqns.~\eqref{eq:BdG}. For a homogeneous superconductor, Eqn.~\eqref{eq:frac} reduces to the standard BCS form~\cite{Tinkham}.}

If the bound states have a different relaxation time than the continuum states, we treat them as two parallel resistive channels with conductivity
\begin{equation}
\sigma_1 = \left(n_b \tau_b + n_c \tau_c \right) e^2 / m,
\label{eq:3fl}
\end{equation}
where $n_b$ and $n_c$ are the densities of bound and continuum states, and $\tau_b$ and $\tau_c$ are their respective relaxation times. {These densities are calculated using Eqn.~\eqref{eq:frac} but restricting the summation over states with ${E_n<\Delta_0}$ or ${E_n\geq \Delta_0}$.} We call this modification the ``three-fluid’’ model. 

A decrease in $\sigma_1$ causes more current to flow through the inductive channel, ultimately reducing $R_\text{s}$. Near the surface, as the field increases the density of bound states increases and the density of continuum states decreases. {Accordingly,} any source of scattering that makes $\tau_b \ll \tau_c$ can cause a decrease in $\sigma_1$ and thus an anti-$Q$ slope in our three-fluid model.

\section{Results}
\label{sec:results}
\begin{figure}
\includegraphics[width=.5\textwidth,left]{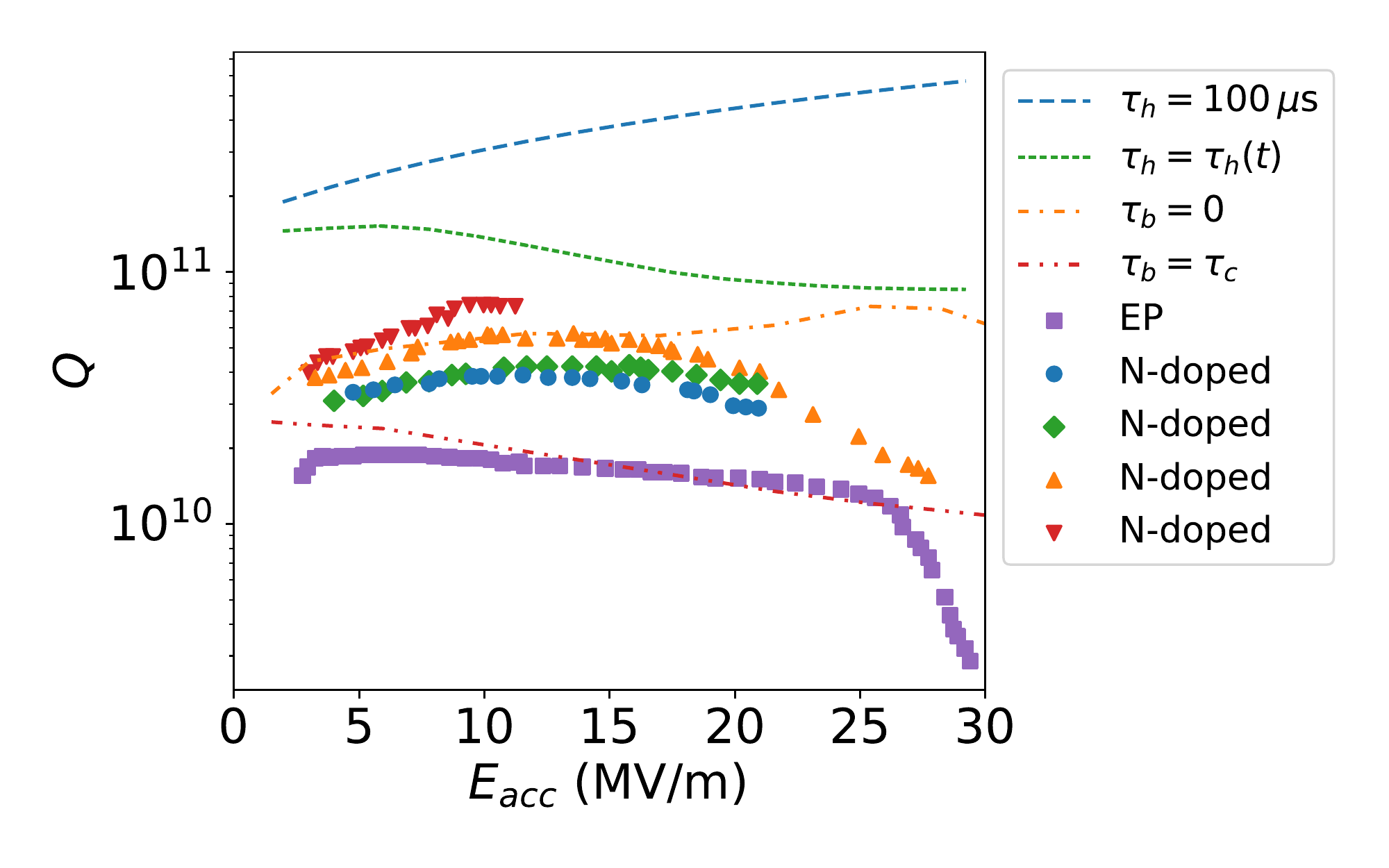}
\caption{Quality factor vs.\ field amplitude for a niobium cavity at frequency $1.3$ GHz. The dashed lines show the quality factor computed from our inelastic scattering disequilibrium mechanism, using either a constant $\tau_h$ or the time-dependent $\tau_h(t)$ in Eqn.~\eqref{eq:tauh}. The dash-dot lines show the quality factor from this mechanism plus the three-fluid model as described in the text {with $\tau_c=1$ ps}. Markers show experimental data from Grassellino et al.\ \cite{Grassellino} for a selection of electro-polished (EP) or nitrogen-doped cavities.}
\label{fig:QvsB}
\end{figure}

Figure~\ref{fig:QvsB} plots the quality factor vs.\ field amplitude for a constant $\tau_h=100$ $\mu$s and for the time-dependent $\tau_h(t)$ in Eqn.~\eqref{eq:tauh}, along with several experimental measurements from Grassellino et al.\ \cite{Grassellino} for comparison. Constant $\tau_h$ yields an anti-$Q$ slope, while the refined $\tau_h$ has $Q$ declining as the field strength increases. Both approximations yield quality factors well above the experimental values, indicating that the mechanism of inelastic scattering into bound states is likely not the dominant source of dissipation, but may become important for cavities with quality factors approaching $10^{11}$. Exactly how important depends in large part on the magnitude of $\tau_h$.

As noted earlier, the parameters for a typical niobium cavity put us in the slow-scattering regime, wherein dissipation would increase if the scattering rates were to increase. A tenfold increase in scattering rates would make the quality factor ten times lower. Given the rough nature of our approximations and the strong influence of $\tau_h$ on the dissipation, more precise calculations of both the magnitude and field dependence of scattering rates are likely warranted.

Another important consideration is the frequency dependence of the quality factor. In the $\omega\tau\gg1$ regime, dissipation is limited by the rarity of scattering events that fill a given bound state. Doubling the frequency leads to half as many scattering events and thus half as much dissipation per cycle, but with twice as many cycles per second the energy dissipated per second remains constant. On the other hand, in the $\omega\tau\ll1$ regime the dissipation is limited by the fact that the bound states stay close to equilibrium. Increasing the frequency leads to more disequilibrium and more cycles per second, so the result is that dissipation is proportional to frequency squared. Numerical computation verifies that dissipation goes with $\omega^2$ when $\omega\tau\ll1$ and flattens out once $\omega\tau\gg1$. Inelastic scattering rates in bulk niobium suggest $\omega\tau\gg1$, so unless the filling of surface states is much faster than what we have inferred from bulk estimates, dissipation from this mechanism is independent of frequency within the typical SRF range.

We also compute the dissipation resulting from the three-fluid model, using our Bogoliubov-de Gennes states to calculate the densities of the two normal fluids. The relative sizes of $\tau_c$ and $\tau_b$ depend on the nature of the scattering. Accurate descriptions of surface scattering will depend on factors such as the cavity's impurity profiles, surface roughness~\cite{Jacob} or coatings such as the Nb$_3$Sn coating often applied to the surface of SRF cavities~\cite{CarlsonPos2021}. When we model the elastic scattering of quasiparticles with point-like impurities using a three-dimensional delta function as the perturbing potential, we estimate using Fermi’s golden rule that $\tau_b$ is shorter than $\tau_c$ for states near the surface by roughly a factor of two. When we model an extended defect such as a surface nanohydride with a one-dimensional delta function $\delta(z-z_0)$, we find values for $\tau_b$ as much as six times smaller than $\tau_c$.

Given the uncertainty about the ratio $\tau_b/\tau_c$, in Figure~\ref{fig:QvsB} we assume $\tau_c=1$ ps \cite{Tinkham} and plot the resulting quality factor for two limiting cases, $\tau_b=0$ (yellow dash-dot curve) and $\tau_b=\tau_c$ (red dash-dot-dot curve), to show the range of plausible $Q$ slopes. The results for $\tau_b=0$ roughly align with the anti-$Q$ slope in the experimental data for the first $15$~MV/m, whereas the $\tau_b=\tau_c$ curve is similar to that of the electropolished cavity with no anti-$Q$ slope. 

As the field strength increases, the density of the bound state fluid near the surface grows while the density of the continuum state fluid shrinks. The increase in bound states exceeds the decrease in continuum states, so if $\tau_b=\tau_c$ the net effect is to increase $\sigma_1$, drawing more current through the resistive channel and dissipating more energy. However, if $\tau_b$ is negligible compared to $\tau_c$, the net effect is to decrease $\sigma_1$, causing less dissipation and yielding an anti-Q slope. We estimate an anti-$Q$ slope to appear approximately when $\tau_b<\tau_c/{5}$. {The magnitude of the quality factor could change significantly depending on the value of $\tau_c$, but the main point is the difference in slope: These results suggest that one could manipulate} the $Q$ slope by modulating any scattering sources that affect the bound states more than the continuum states. 

It is worth noting that in the two- and three-fluid models, as in the conventional linear response theory, the dissipation has an $\omega^2$ dependence. If a mechanism with such dependence produces an anti-$Q$ slope, the addition of a mechanism with flat frequency dependence and no anti-$Q$ slope---such as our inelastic scattering disequilibrium mechanism---would explain why the anti-$Q$ slope becomes more pronounced at higher frequencies. While this argument is speculative, it nonetheless demonstrates the potential for bound states to contribute to the anti-$Q$ slope.

\section{Conclusion}
\label{sec:conclusion}
For superconductors in large AC fields there are quasiparticle states for which a linear response approach to dissipation is inadequate. These states are bound to the surface with energies below the value of the bulk superconducting gap, and their energies change throughout an AC cycle. By focusing on these fundamentally nonperturbative bound states we have made a stark departure from the conventional theories derived in a weak field setting. For the range of parameters we expect in a typical niobium SRF cavity, the dissipation from inelastic quasiparticle-phonon scattering into these states is independent of frequency and is likely not the dominant source of dissipation. However, if inelastic scattering is much stronger at the surface of the superconductor than it is in bulk, the dissipation from this mechanism could become important. More refined calculations of scattering rates are a sensible next step to build on these results. 

Moreover, we argue that when the relaxation time of the bound states differs from that of the continuum states, the two-fluid model should be modified to consider the continuum and bound quasiparticles as two separate normal-conducting fluids in addition to the superconducting fluid. Depending on whether the relaxation time of the bound state fluid is comparable to or much smaller than that of the continuum state fluid, the resulting quality factor could either increase or decrease with field strength. Our estimations suggest that some kinds of scattering can indeed affect the bound states more than the continuum states. If one can identify and control the concentration of such scattering sources in an SRF cavity, in principle one could use them to activate the anti-$Q$ slope.

\section{Acknowledgments}
This work was supported by the U.S. National Science Foundation under Award PHY-1549132, the Center for Bright Beams.

\bibliographystyle{apsrev4-2}
\bibliography{BdG}

\end{document}